\documentclass[a4paper,12pt]{article}
\usepackage{amsmath}
\usepackage{amssymb}
\usepackage{amsfonts}
\usepackage{amssymb,amsmath}
\usepackage{cancel}
\usepackage[dvips]{lscape,graphicx,epsfig}
\usepackage{fleqn}
\usepackage[footnotesize]{caption}
\usepackage{mathrsfs}

\def\beq{\begin{equation}}
\def\eeq{\end{equation}}
\def\bea{\begin{eqnarray}}
\def\eea{\end{eqnarray}}

\def\<{\left\langle}
\def\>{\right\rangle}

\renewcommand{\baselinestretch}{1.30}

\newcommand{\bc}{\begin{center}}
\newcommand{\ec}{\end{center}}
\newcommand{\bd}{\begin{displaymath}}
\newcommand{\ed}{\end{displaymath}}
\newcommand{\be}{\begin{equation}}
\newcommand{\ee}{\end{equation}}
\newcommand{\ba}{\begin{array}}
\newcommand{\ea}{\end{array}}
\newcommand{\bt}{\begin{tabular}}
\newcommand{\et}{\end{tabular}}

\newcommand{\ds}{\displaystyle}

\begin{document}

\bibliographystyle{OurBibTeX}

\begin{titlepage}

 \vspace*{-15mm}
\begin{flushright}
{SHEP--06--39}\\
\today\\
\end{flushright}
\vspace*{5mm}

\begin{center}
{
\sffamily
\Large Gauge Coupling Unification in the Exceptional Supersymmetric
Standard Model}
\\[8mm]
S.F.~King$^a$\footnote{E-mail: \texttt{sfk@hep.phys.soton.ac.uk}.},
S.~Moretti$^a$\footnote{E-mail: \texttt{stefano@hep.phys.soton.ac.uk}.},
R.~Nevzorov$^b$\footnote{On leave of absence from the Theory Department, ITEP, Moscow, Russia;\\[-2mm]
$~~~~~$ E-mail: \texttt{r.nevzorov@physics.gla.ac.uk}}
\\[3mm]
{\small\it
$^a$ School of Physics and Astronomy, University of Southampton,\\
Southampton, SO17 1BJ, U.K.\\[2mm]
$^b$ Department of Physics and Astronomy, University of Glasgow,\\
Glasgow G12 8QQ, U.K.
}\\[1mm]
\end{center}
\vspace*{0.75cm}

\begin{abstract}
\noindent
We consider the renormalisation group flow of gauge couplings within
the so-called exceptional supersymmetric standard model
(E$_6$SSM) based
on the low energy matter content of 27 dimensional representations of
the gauge group $E_6$, together
with two additional non-Higgs doublets.
The two--loop beta functions are computed, and the threshold
corrections are studied in the E$_6$SSM.
Our results show that
gauge coupling unification in the E$_6$SSM can be achieved for
phenomenologically acceptable values of $\alpha_3(M_Z)$,
consistent with the central measured low energy value,
unlike in the minimal supersymmetric standard model (MSSM)
which, ignoring the
effects of high energy threshold corrections, requires
significantly higher values of $\alpha_3(M_Z)$, well above the
experimentally measured central value.
\end{abstract}

\end{titlepage}
\newpage
\setcounter{footnote}{0}

\section{Introduction}

Unification of gauge couplings is probably one of the most appealing
features of supersymmetric (SUSY) extensions of the standard model
(SM).  More than fifteen years ago it was found that the electroweak
(EW) and strong gauge couplings extracted from LEP data (hence at the
EW scale) and extrapolated to high energies using the renormalisation
group equation (RGE) evolution do not meet within the SM but converge
to a common value at some high energy scale (within $\alpha_3(M_Z)$
uncertainties) after the inclusion of supersymmetry, e.g. in the
framework of the minimal SUSY standard model (MSSM) \cite{1}. This
allows one to embed SUSY extensions of the SM into Grand Unified
Theories (GUTs) (and superstring ones) that make possible partial
unification of gauge interactions with gravity.  Simultaneously, the
incorporation of weak and strong gauge interactions within GUTs
permits to explain the peculiar assignment of $U(1)_Y$ charges
postulated in the SM and to address the observed mass hierarchy of
quarks and leptons.

Due to the lack of direct evidence verifying or falsifying the
presence of superparticles at low energies, gauge coupling unification
remains the main motivation for low--energy supersymmetry based on
experimental data.  But since 1990 the uncertainty in the
determination of $\alpha_3(M_Z)$ has reduced significantly and the
analysis of the two--loop RG flow of gauge couplings performed in
\cite{21}--\cite{2} revealed that it is rather problematic to achieve
exact unification of gauge couplings within the MSSM. This is also
demonstrated in Fig.~1a, where we plot the running of the gauge
couplings from the EW ($M_Z$) scale to the GUT ($M_X$) scale.
Fig.~1b shows a blow--up of the crucial region in the vicinity of 
$M_X=3\cdot 10^{16}\,\mbox{GeV}$. To ensure the correct breakdown 
of the EW symmetry requires an effective SUSY threshold scale around 
$250\,\mbox{GeV}$, which corresponds to a SUSY Higgs mass parameter 
$\mu\simeq 1.5\,\mbox{TeV}$. Dotted lines show the interval of
variations of gauge couplings caused by $1\,\sigma$ deviations of
$\alpha_3(M_Z)$ around its average value, i.e.  $\alpha_3(M_Z)\simeq
0.118\pm 0.002$ \cite{3}. From Fig.~1b it is clear that exact
gauge coupling unification in the MSSM cannot be attained even within
$2\,\sigma$ deviations from the current average value of
$\alpha_3(M_Z)$. Recently, it was argued that it is possible to get 
the unification of gauge couplings in the minimal SUSY model for 
$\alpha_3(M_Z)=0.123$ \cite{31}.

The above observation is in fact true for a whole class of GUTs that
break to the SM gauge group in one step and which predict a so--called
``grand desert" between the EW and GUT scales. This conclusion 
must be qualified, however, by the fact that in general there are 
non--negligible high energy 
GUT/string threshold corrections to the running of the
couplings associated with heavy particle thresholds and higher
dimension operator effects which we shall not consider here.  
Furthermore, in this paper, we restrict our
considerations to the minimal scenario for GUT symmetry group
breakdown --- the aforementioned one--step GUTs -- as this allows one
to get a stringent prediction for $\alpha_s(M_Z)$. In particular, we
examine gauge coupling unification within an $E_6$ inspired extension
of the MSSM, the exceptional supersymmetric standard model (E$_6$SSM) of
Refs.~\cite{4}--\cite{5} in which the $E_6$ symmetry breaking proceeds 
uniquely at a single step through the $SU(5)$ breaking direction.
This results in a low energy SM gauge group
augmented by a unique $U(1)_{N}$ gauge group under which
right-handed neutrinos have zero charge, allowing them to be
superheavy, shedding light on
the origin of the mass hierarchy in the lepton sector and providing a
mechanism for the generation of the lepton and baryon asymmetry of the
Universe. The $\mu$ problem of the MSSM is solved within the E$_6$SSM
in a similar way to the NMSSM, but without the accompanying problems
of singlet tadpoles or domain walls. Thus the E$_6$SSM is a low energy
alternative to the MSSM or NMSSM.

In this paper
we calculate the two--loop beta functions of the
gauge couplings in the E$_6$SSM, and then apply them to
the question of gauge coupling unification, 
including the important effects of low energy threshold corrections.
The structure of the two--loop
contributions to the corresponding beta functions is such that the EW
and strong couplings meet at some high energy scale for an
$\alpha_3(M_Z)$ value which is just slightly higher than the 
experimentally measured central value, with the low energy threshold
effects pushing it further towards the central measured value.
As the results in Fig.~1c,d will show, 
the unification of gauge couplings in the E$_6$SSM is
achieved for values of $\alpha_3(M_Z)$ consistent with the measured
central value, unlike in the MSSM which, ignoring the
effects of high energy threshold corrections, requires
significantly higher values of $\alpha_3(M_Z)$, well above the
experimentally measured central value.

The layout of the remainder of the paper is as follows.
In section 2 we present an analytical approach to the
solution of the RGEs for the gauge couplings that allows one to
examine the unification of forces in SUSY models and we specialise to
the MSSM case in section 3. In section 4 we briefly review the
E$_6$SSM and in section 5 we discuss the two--loop RG
flow of the gauge couplings within this model, including the
low energy threshold corrections, leading to the stated results.
Section 6 concludes the paper.

\section{RG flow of gauge couplings in SUSY models}

In SUSY models the running of the SM gauge couplings is described 
by a system of RGEs which can be written in the following form:
\be
\frac{d \alpha_i}{dt}=\ds\frac{\beta_i \alpha_i^2}{(2\pi)}\,,
\qquad\qquad\beta_i=b_i+\ds\frac{\tilde{b}_i}{4\pi}\,, 
\label{gc0}
\ee
where $b_i$ and $\tilde{b}_i$ are one--loop and two--loop contributions to the 
beta functions \cite{6}--\cite{7}, $t=\ln\left(\mu/M_Z\right)$, $\mu$ is a renormalisation scale, with the
index $i$ running from $1$ to $3$ corresponding to $U(1)_Y$, $SU(2)_W$ and $SU(3)_C$ interactions,
respectively. One can obtain an approximate solution of the RGEs in Eq.~(\ref{gc0})
that at high energies can be written as \cite{Chankowski:1995dm}
\be
\ds\frac{1}{\alpha_i(t)}=\frac{1}{\alpha_i(M_Z)}-\ds\frac{b_i}{2\pi} t-\frac{C_i}{12\pi}-\Theta_i(t)
+\ds\frac{b_i-b_i^{SM}}{2\pi}\ln\frac{T_i}{M_Z}\,,
\label{gc1}
\ee
where the third term in the right--hand side of Eq.~(\ref{gc1}) is the
$\overline{MS}\to\overline{DR}$ conversion factor with $C_1=0$, $C_2=2$, $C_3=3$ \cite{8},
\be
\Theta_i(t)=\ds\frac{1}{8\pi^2}\int_0^t \tilde{b}_i d\tau\,,\qquad\qquad
T_i=\prod_{k=1}^N\biggl(m_k\biggr)^{\ds\frac{\Delta b^k_i}{b_i-b_i^{SM}}}\,
\label{gc2}
\ee 
and $b_i^{SM}$ are the coefficients of the
one--loop beta functions in the SM,
while $m_k$ and $\Delta b_i^k $ are masses and contributions to the beta functions due
to new particles appearing in the considered SUSY models. Because the 
two--loop corrections to the running of the gauge couplings $\Theta_i(t)$ are considerably 
smaller than the leading terms,  the gauge and Yukawa couplings in $\tilde{b}_i$
are usually replaced by the corresponding solutions of the RGEs obtained in 
the one--loop approximation. The threshold corrections associated with the last terms 
in Eq.~(\ref{gc1}) are of the same order as or even less than $\Theta_i(t)$. 
Therefore in Eqs.~(\ref{gc1})--(\ref{gc2}) only leading one--loop threshold effects are 
taken into account.

Relying on the approximate solution of the RGEs in
Eqs.~(\ref{gc1})--(\ref{gc2}) one can 
establish the relationships between the values of the gauge couplings at the EW 
and GUT scales, for any general SUSY model. Then by using the expressions describing the RG flow 
of $\alpha_1(t)$ and $\alpha_2(t)$ it is rather easy to find the scale $M_X$ where 
$\alpha_1(M_X)=\alpha_2(M_X)=\alpha_0$ and the value of the overall gauge coupling 
$\alpha_0$ at this scale. Substituting $M_X$ and $\alpha_0$ into the solution of 
the RGE for the strong gauge coupling one finds the value of $\alpha_3(M_Z)$ for 
which exact gauge coupling unification takes place \cite{Carena:1993ag}:
\be
\ba{c}
\ds\frac{1}{\alpha_3(M_Z)}=\frac{1}{b_1-b_2}\biggl[\ds\frac{b_1-b_3}{\alpha_2(M_Z)}-
\ds\frac{b_2-b_3}{\alpha_1(M_Z)}\biggr]-\frac{1}{28\pi}+\Theta_s-\Delta_s\,,\\[5mm]
\Theta_s=\biggl(\ds\frac{b_2-b_3}{b_1-b_2}\Theta_1-\frac{b_1-b_3}{b_1-b_2}\Theta_2+\Theta_3\biggr)\,,
\qquad \Theta_i=\Theta_i(M_X),
\ea
\label{gc4}
\ee
where $\Delta_s$ are combined threshold corrections whose precise
form depends on the model under consideration.

\section{MSSM}
In this section we apply the results of the previous section to the MSSM.
In the MSSM $\Delta_s$ takes the form 
\cite{21}--\cite{Chankowski:1995dm}, \cite{Carena:1993ag}--\cite{Langacker:1992rq}:
\be
\Delta_s=-\frac{19}{28\pi}\ln\frac{M_{S}}{M_Z}\,,\qquad\qquad 
M_S=\ds\frac{T_2^{100/19}}{T_1^{25/19}T_3^{56/19}}\,.
\label{gc5}
\ee   
For example, assuming for simplicity that superpartners of all quarks
are degenerate, i.e. their masses are 
equal to $m_{\tilde{q}}$, and all sleptons have a common mass
$m_{\tilde{l}}$, we find:
\be
M_S\simeq \mu\biggl(\ds\frac{m_A}{\mu}\biggr)^{3/19}\biggl(\ds\frac{M_2}{\mu}\biggr)^{4/19}
\biggl(\frac{M_2}{M_3}\biggr)^{28/19}\biggl(\ds\frac{m_{\tilde{l}}}{m_{\tilde{q}}}\biggr)^{3/19}\,.
\label{gc6}
\ee
In Eq.~(\ref{gc6}) $M_3$ and $M_2$ are masses of gluinos and winos 
(superpartners of $SU(2)_W$ gauge bosons), whereas $\mu$ and $m_A$ are
$\mu$--term and masses of heavy Higgs states respectively.   
In general $T_1$, $T_2$ and $T_3$, obtained from Eq.~(\ref{gc2}),
can be quite different. 

We now perform a simplified numerical discussion of the previous results
in order to illustrate 
the effect of threshold corrections on gauge unification in the MSSM.  In
our simplified discussion we shall assume the effective
threshold scales $T_i$ be equal to each other, $T_1=T_2=T_3=M_S$,
where the last equality follows from Eq.~(\ref{gc5}). 
From Eqs.~(\ref{gc4})--(\ref{gc5}) and Tab.~\ref{gc} it follows that,
in order to achieve gauge coupling unification in the MSSM
with $\alpha_s(M_Z)\simeq 0.118$, the effective threshold scale must
be around $M_S\approx 1\,\mbox{TeV}$. However the correct pattern of EW
symmetry breaking (EWSB) requires $\mu$ to lie within the $1-2\,\mbox{TeV}$
range, while from Eq.~(\ref{gc6}) it follows that $M_S\simeq \mu/6$,
which implies that the
effective threshold scale should be $M_S<200-300\,\mbox{GeV}$
\cite{21}--\cite{Chankowski:1995dm}, \cite{Carena:1993ag}--\cite{9}. 
For such small values of the scale $M_S$ exact gauge coupling unification can
be obtained only for large values of $\alpha_3(M_Z)\gtrsim 0.123$,
which are disfavoured by the recent fit to experimental data.
Put it another way, assuming that the low energy QCD coupling is 
at its central value $\alpha_s(M_Z)=0.118$, and assuming 
$M_S=250$ GeV, the gauge couplings fail to meet
exactly at the GUT scale, as shown in Fig.~1a--1b.
As we shall show, this situation is improved dramatically in the 
E$_6$SSM.

\section{E$_6$SSM - a brief review}
In this section, in order to make the paper
self-contained, we give a brief review of the E$_6$SSM
which was proposed recently in \cite{4,5}.
The E$_6$SSM involves an additional low energy gauged 
$U(1)_N$ not present in the MSSM, and in order
to ensure anomaly cancellation the
particle content of the E$_6$SSM is also extended to include
three complete fundamental
$27$ representations of $E_6$ at low energies.
These multiplets decompose under the
$SU(5)\times U(1)_{N}$ subgroup of $E_6$ as follows \cite{10}:
\begin{equation}
\begin{array}{c}
27_i\to \left(10,\,\displaystyle\frac{1}{\sqrt{40}}\right)_i+
\left(5^{*},\,\displaystyle\frac{2}{\sqrt{40}}\right)_i
+\left(5^{*},\,-\displaystyle\frac{3}{\sqrt{40}}\right)_i +
\left(5,-\displaystyle\frac{2}{\sqrt{40}}\right)_i+\\[3mm]
+\left(1,\displaystyle\frac{5}{\sqrt{40}}\right)_i+\left(1,0\right)_i.
\end{array}
\label{essm2}
\end{equation}
The first and second quantities in the brackets are the $SU(5)$
representation and extra $U(1)_{N}$ charge while $i$ is a family index
that runs from 1 to 3. An ordinary SM family which contains the
doublets of left-handed quarks $Q_i$ and leptons $L_i$, right-handed
up- and down-quarks ($u^c_i$ and $d^c_i$) as well as right-handed
charged leptons, is assigned to $\left(10,\frac{1}{\sqrt{40}}\right)_i
+\left(5^{*},\,\frac{2}{\sqrt{40}}\right)_i$. Right-handed neutrinos
$N^c_i$ should be associated with the last term in Eq.~(\ref{essm2}),
$\left(1,0\right)_i$.  The next-to-last term in Eq.~(\ref{essm2}),
$\left(1,\frac{5}{\sqrt{40}}\right)_i$, represents SM-type singlet fields
$S_i$ which carry non-zero $U(1)_{N}$ charges and therefore survive
down to the EW scale.  The pair of $SU(2)_W$-doublets ($H_{1i}$ and $H_{2i}$)
that are contained in $\left(5^{*},\,-\frac{3}{\sqrt{40}}\right)_i$ and
$\left(5,-\frac{2}{\sqrt{40}}\right)_i$ have the quantum numbers of
Higgs doublets. So they form either Higgs or non--Higgs $SU(2)_W$ multiplets.
Other components of these $SU(5)$ multiplets form colour
triplets of exotic quarks $\overline{D}_i$ and $D_i$ with electric
charges $-1/3$ and $+1/3$ respectively. However these exotic quark states
carry a $B-L$ charge $\left(\pm\frac{2}{3}\right)$ twice larger than that of
ordinary ones. Therefore in phenomenologically viable $E_6$ inspired models
they can be either diquarks or leptoquarks. In addition to the complete $27_i$
multiplets the low energy particle spectrum of the E$_6$SSM is supplemented by
$SU(2)_W$ doublet $H'$ and anti-doublet $\overline{H}'$ states from 
the extra $27'$ and 
$\overline{27'}$ to preserve gauge coupling unification. Thus, in addition to a 
$Z'$ corresponding to the $U(1)_N$ symmetry, the E$_6$SSM involves extra matter
beyond the MSSM that forms three $5+5^{*}$ representations of $SU(5)$
plus three $SU(5)$ singlets with $U(1)_N$ charges. The presence of a $Z'$ boson 
and exotic quarks predicted by the E$_6$SSM provides spectacular new physics 
signals at the LHC which were discussed in \cite{4}--\cite{5}, \cite{11}.

The superpotential in $E_6$ inspired models involves a lot of new
Yukawa couplings in comparison to the SM. In general these new
interactions induce non--diagonal flavour transitions. To avoid a
flavour changing neutral current (FCNC) problem an extra $Z^{H}_2$ symmetry 
is postulated in the E$_6$SSM. Under this symmetry all superfields 
except one pair of $H_{1i}$ and $H_{2i}$ (say $H_d\equiv H_{13}$ and 
$H_u\equiv H_{23}$) and one SM-type singlet field ($S\equiv S_3$) are odd.
The $Z^{H}_2$ symmetry reduces the structure of the Yukawa interactions to:
\begin{equation}
\begin{array}{rcl}
W_{\rm ESSM}&\simeq & \lambda_i S(H_{1i}H_{2i})+\kappa_i
S(D_i\overline{D}_i)+f_{\alpha\beta}S_{\alpha}(H_d
H_{2\beta})+ \tilde{f}_{\alpha\beta}S_{\alpha}(H_{1\beta}H_u)+\\[2mm]
&&+\mu'(H'\overline{H}')+g_{i}e^c_i(H_d H')+W_{\rm{MSSM}}(\mu=0),
\end{array}
\label{essm3}
\end{equation}
where $\alpha,\beta=1,2$ and $i=1,2,3$\,. The $SU(2)_W$ doublets $H_u$ 
and $H_d$ play the role of Higgs fields generating the masses of 
quarks and leptons after EWSB. Therefore it is 
natural to assume that only $S$, $H_u$ and $H_d$ acquire non-zero vacuum 
expectation values (VEVs). The VEV of the SM-type
 singlet field $S$ breaks the extra 
$U(1)_N$ symmetry thereby providing an effective $\mu$ term as well as 
the necessary exotic fermion masses and also inducing that of the $Z'$ boson. 
To guarantee that only $H_u$, $H_d$ and $S$ develop VEVs in the E$_6$SSM 
a certain hierarchy between the Yukawa couplings is imposed, i.e. 
$\lambda_3\gtrsim \lambda_{1,2}\gg f_{\alpha\beta}, \tilde{f}_{\alpha\beta}, g_{i}$.

However the $Z^{H}_2$ symmetry can only be an approximate one because it forbids 
all Yukawa interactions that would allow the exotic quarks to decay. Since 
models with stable charged exotic particles are ruled out by different 
experiments \cite{12} the $Z^{H}_2$ symmetry has to be broken. 
At the same time the breakdown of $Z^{H}_2$ should not give rise to  
operators leading to rapid proton decay. There are two ways to overcome 
this problem. The resulting Lagrangian has to be invariant with 
respect to either a $Z_2^L$ symmetry, under which all superfields except lepton ones 
are even, or a $Z_2^B$ discrete symmetry, which implies that 
exotic quark and lepton superfields are odd whereas the others remain even. 
Because $Z^{H}_2$ symmetry violating operators may also give an appreciable 
contribution to the amplitude of $K^0-\overline{K}^0$ oscillations and 
give rise to new muon decay channels like $\mu\to e^{-}e^{+}e^{-}$
the corresponding Yukawa couplings are expected to be small. 
Therefore $Z^{H}_2$ symmetry violating Yukawa couplings are irrelevant
for the analysis of the RG flow of gauge couplings considered here.

It is worth to emphasize that all the discrete symmetries $Z^{H}_2$, $Z_2^L$ and $Z_2^B$ 
that we use here to prevent rapid proton decay break $E_6$ because different components 
of the fundamental $27$ representation transform differently under these symmetries.
Another manifestation of the breakdown of the $E_6$ symmetry is the presence
of the $SU(2)_W$ doublet $H'$ and anti-doublet $\overline{H}'$ in the low energy particle
spectrum of the E$_6$SSM that comes from the splitting of extra $27'$ and $\overline{27}'$.
Because the splitting of $27$--plets is a necessary ingredient of the considered model, as it
 is required in order to attain gauge coupling unification, it seems to be very attractive 
to reduce all origins of the $E_6$ symmetry breakdown (including postulated discrete 
symmetries) to the splitting of different $E_6$ multiplets. The splitting of GUT 
multiplets can be naturally achieved in the framework of orbifold GUTs \cite{121}.

The $E_6$ GUT model whose incomplete multiplets form the particle content
of the E$_6$SSM at low energies involves at least eight 27 and one
$\overline{27}$ multiplets. One 27--plet $\Phi_0$ includes only five
components that survive down to the EW scale and compose the Higgs
sector of the E$_6$SSM, namely $S,H_u,H_d$. 
Such $E_6$ GUT model should also
have three pairs of 27--plets $\Phi_i$ and $\Phi^L_i$ which
accommodate three generations of quarks and leptons, where $i$ is a
family index. The $E_6$ multiplets $\Phi^L_i$ contain left-handed and
right-handed lepton superfields ($L_i, e^c_i, N^c_i$) while the $\Phi_i$'s
involve all quark superfields ($Q_i, u^c_i, d^c_i$) as well as
non--Higgs and SM singlet fields. The only exception is $\Phi_3$, that
does not include either non--Higgs or SM-type singlet fields. Exotic
quarks $\overline{D}_i$ and $D_i$ belong either to $\Phi^L_i$ (if they
are leptoquarks) or to $\Phi_i$ (if they are diquarks).  Finally
extra $27'$ and $\overline{27}'$ ($\Phi'$ and $\overline{\Phi}'$)
contain only two light components each that form the $SU(2)_W$ doublet
$H'$ and anti-doublet $\overline{H}'$ with quantum numbers of
left-handed lepton fields.

\begin{table}[ht]
  \centering
  \begin{tabular}{|c|c|c|c|c|c|c|}
    \hline
           & $\Phi_0$ & $\Phi_i$ & $\Phi_i^L$ & $\Phi'$ & $\overline{\Phi}'$ & $\Sigma$ \\
 \hline
$Z^{H}_2$  & $+$      & $-$      & $-$        & $-$     & $-$                & $-$ \\
 \hline
$Z'_2$     & $+$      & $+$      & $-$        & $-$     & $-$                & $+$ \\
 \hline
  \end{tabular}
  \caption{Transformation properties of $E_6$ multiplets
under $Z^H_2$ and $Z'_2$ discrete symmetries.}
  \label{z2}
\end{table}

In order to get a suitable pattern of Yukawa couplings postulated
above we impose the invariance of the Lagrangian of the considered
$E_6$ GUT model under the $Z^H_2\otimes Z'_2$ symmetry. As before all
$E_6$ multiplets except $\Phi_0$ are odd under the $Z^{H}_2$ symmetry
transformations. The $Z'_2$ symmetry is equivalent to either the $Z_2^L$
symmetry when exotic quarks are diquarks or the $Z_2^B$ symmetry if exotic
quarks are leptoquarks. The transformation properties of $E_6$
multiplets under the $Z^H_2$ and $Z'_2$ symmetries are summarised in
Table~\ref{z2}. Here we also introduce the singlet field $\Sigma$ that does
not participate in the $E_6$ gauge interactions. Just as other $E_6$
multiplets, $\Sigma$ is odd under $Z^{H}_2$.

The most general superpotential which is invariant under $E_6$ and $Z^H_2\otimes Z'_2$
symmetry transformations is given by
\begin{equation}
\begin{array}{c}
W=\lambda \Phi_0^3+ \sigma_{ij}\Phi_0 \Phi_i \Phi_j+ \widetilde{\sigma}_{ij}\Phi_0 \Phi_i^L \Phi_j^L
+\ds\frac{\Sigma}{M_{Pl}}\biggl(\eta_i\Phi_0^2\Phi_i+\zeta_{ijk}\Phi_i\Phi_j\Phi_k+\\[3mm]
+\widetilde{\zeta}_{ijk}\Phi_i\Phi^L_j\Phi^L_k\biggr)+\mu_X\Sigma^2+\xi\ds\frac{\Sigma^4}{M_{Pl}}+...\,\,\, .
\end{array}
\label{essm5}
\end{equation}
In the superpotential (\ref{essm5}) we omit higher order terms that
are suppressed as $1/M_{Pl}^2$ or even stronger. If $\mu_X<< M_{Pl}$
the singlet field $\Sigma$ may acquire vacuum expectation value which
is many orders of magnitude smaller than the Planck scale.  Non--zero
vacuum expectation value of $\Sigma$ breaks the $Z_2^H$ symmetry
spontaneously.  Then the first three terms in Eq.~(\ref{essm5}) result
in the $Z_2^H$ symmetric part of the superpotential of the E$_6$SSM at low
energies while the next three terms give rise to couplings that violate
the $Z_2^H$ symmetry explicitly. In this case the effective Yukawa
couplings which are induced after the breakdown of the $Z_2^H$ symmetry
are naturally suppressed by the small ratio
$\ds\frac{<\Sigma>}{M_{Pl}}$ leading to the desirable hierarchical
structure of Yukawa interactions postulated in the E$_6$SSM.

\section{Gauge Coupling Unification in the E$_6$SSM}
We now turn to the central issue of this paper, that of gauge coupling
unification in the E$_6$SSM. We first present our results for 
the two-loop beta functions in this model, before going on to consider
the question of gauge coupling unification in the presence of
low energy threshold effects.
The running of gauge couplings in the E$_6$SSM is affected by a kinetic 
term mixing \cite{4}, \cite{13}. As a result the RGEs can be written 
as follows:
\be
\ds\frac{d G}{d t}=G\times B\,,\qquad\qquad 
\frac{d g_2}{dt}=\ds\frac{\beta_2 g_2^3}{(4\pi)^2}\,,\qquad\qquad
\frac{d g_3}{dt}=\frac{\beta_3 g_3^3}{(4\pi)^2}\,,
\label{gc7}
\ee
where $B$ and $G$ are $2\times 2$ matrices 
\be
G=\left(
\ba{cc}
g_1 & g_{11}\\[2mm]
0   & g'_1
\ea
\right)\,,\qquad
B=\ds\frac{1}{(4\pi)^2}
\left(
\ba{cc}
\beta_1 g_1^2 & 2g_1g'_1\beta_{11}+2g_1g_{11}\beta_1\\[2mm]
0 & g^{'2}_1\beta'_1+2g'_1 g_{11}\beta_{11}+g_{11}^2\beta_1
\ea
\right)\,.
\label{gc8}
\ee
As always the two--loop diagonal $\beta_i$ and off--diagonal $\beta_{11}$ beta functions
may be presented as a sum of one--loop and two--loop contributions (see Eq.~(\ref{gc0})). In the one--loop
approximation the beta functions are given by
\be
\ba{rclrcl}
b_1&=&\ds\frac{3}{5}+3N_g\,,\qquad &
b'_1&=&\ds\frac{2}{5}+3N_g\,,\qquad
b_{11}=\ds\frac{\sqrt{6}}{5}\,,\\[2mm]
b_2&=&-5+3N_g\,,\qquad &
b_3&=&-9+3N_g\,.
\ea
\label{gc10}
\ee
The parameter $N_g$ appearing~ in Eq.~(\ref{gc10}) is the number of generations in the E$_6$SSM 
forming complete $E_6$ fundamental representations at low energies ($E<<M_X$). As one can easily see
$N_g=3$ is the critical value for the one--loop beta function of the
strong interactions.
Since by construction three complete 27--plets survive to low energies in the E$_6$SSM 
$\tilde{b}_3$ is equal to zero in our case and in the one--loop approximation the 
$SU(3)_C$ gauge coupling remains constant everywhere from $M_Z$ to $M_X$. Because of this 
any reliable analysis of gauge coupling unification requires the inclusion of
two--loop corrections to the beta functions of gauge couplings. Here we calculate 
the two--loop contributions to the diagonal beta functions only. Using the results
of the computation of two--loop beta functions in a general softly broken $N=1$ SUSY model \cite{7}
we find the following two--loop beta functions for the E$_6$SSM:
\be
\ba{rcl}
\tilde{b}_1&=& 8N_g \alpha_3+\left(\ds\frac{9}{5}+3N_g\right)\alpha_2+\left(\ds\frac{9}{25}+3 N_g\right) \alpha_1+
\left(\ds\frac{6}{25}+N_g\right) \alpha'_1-\\[3mm]
&&-\ds\frac{26}{5} y_t-\ds\frac{14}{5} y_b-\ds\frac{18}{5} y_{\tau}-\ds\frac{6}{5}\Sigma_{\lambda}-
\ds\frac{4}{5}\Sigma_{\kappa}\,,\\[3mm] 
\tilde{b}'_1&=& 8N_g \alpha_3+\left(\ds\frac{6}{5}+3N_g\right)\alpha_2+
\left(\ds\frac{6}{25}+ N_g\right) \alpha_1+\left(\ds\frac{4}{25}+3N_g\right) \alpha'_1-\\[3mm]
&&-\ds\frac{9}{5} y_t-\ds\frac{21}{5}y_b-\ds\frac{7}{5}y_{\tau}-\ds\frac{19}{5}\Sigma_{\lambda}-
\ds\frac{57}{10}\Sigma_{\kappa}\,,\\[3mm]
\tilde{b}_2&=&8N_g \alpha_3+\biggl(-17+21 N_g\biggr)\alpha_2+ \left(\ds\frac{3}{5}+N_g\right) \alpha_1+
\left(\ds\frac{2}{5}+N_g\right) \alpha'_1-\\[3mm]
&&-6 y_t-6 y_b-2 y_{\tau}-2\Sigma_{\lambda}\,,\\[3mm]
\tilde{b}_3&=&\alpha_3\biggl(-54+34 N_g\biggr)+3 N_g \alpha_2+ N_g \alpha_1+N_g \alpha'_1-4 y_t-4 y_b- 2\Sigma_{\kappa}\,,\\[3mm]
\Sigma_{\lambda}&=&y_{\lambda_1}+y_{\lambda_2}+y_{\lambda_3}\,,\qquad\qquad\qquad\Sigma_{\kappa}=y_{\kappa_1}+y_{\kappa_2}+y_{\kappa_3}\,,
\ea
\label{gc11}
\ee
where $y_t=\ds\frac{h_t^2}{4\pi}$,\, $y_b=\ds\frac{h_b^2}{4\pi}$,\, $y_{\tau}=\ds\frac{h_{\tau}^2}{4\pi}$,\, 
$y_{\lambda_i}=\ds\frac{\lambda_i^2}{4\pi}$ and $y_{\kappa_i}=\ds\frac{\kappa_i^2}{4\pi}$.
Because our previous analysis performed in \cite{4} revealed that an off--diagonal gauge coupling $g_{11}$ being set to zero at 
the scale $M_X$ remains very small at any other scale below $M_X$ we neglect two--loop corrections to the 
off--diagonal beta function $\beta_{11}$. 

The results of our numerical studies of gauge coupling unification
in this model are summarised in Fig.~1c--d
where the two--loop RG flow of gauge couplings in the E$_6$SSM is
shown. As before we fix the effective SUSY threshold scale to be equal
to $250\,\mbox{GeV}$, that on the one hand results in appreciable
threshold corrections to the RG running of the gauge couplings but on the
other hand does not spoil the breakdown of the EW symmetry. We
also assume that the masses of the $Z'$ and all exotic fermions and bosons
predicted by the E$_6$SSM are degenerate around $1.5\,\mbox{TeV}$. Thus we
use the SM beta functions to describe the running of gauge couplings
between $M_Z$ and $M_S$, then we apply the two--loop RGEs of the
MSSM to compute the flow of $g_i(t)$ from $M_S$ to $M_{Z'}$ and the
two--loop RGEs of the E$_6$SSM to calculate the evolution of
$g_i(t)$ between $M_{Z'}$ and $M_X$ which is equal to 
$3.5\cdot 10^{16}\,\mbox{GeV}$ in the case of the E$_6$SSM. Again dotted 
lines in Fig.~1c--d represent the changes of the evolution of gauge
couplings induced by the variations of $\alpha_3(M_Z)$ within
$1\,\sigma$ around its average value. From Fig.~1a--d one
can easily see that the interval of variations of $\alpha_3(t)$ is
always considerably wider than the ones for $\alpha_1(t)$ and
$\alpha_2(t)$. However one may expect that the dependence of
$\alpha_1(t)$ and $\alpha_2(t)$ on the value of the strong gauge coupling
at the EW scale should be relatively weak because
$\alpha_3(t)$ appears only in the two--loop contributions to the
corresponding beta functions.  

It is also worth to notice that at high
energies the uncertainty in $\alpha_3(t)$ caused by the variations of
$\alpha_3(M_Z)$ is much bigger in the E$_6$SSM than in the MSSM. This
happens because in the MSSM the strong gauge coupling reduces with
increasing renormalisation scale. Therefore the interval of
variations of $\alpha_3(t)$ near the scale $M_X$ shrinks
drastically. This focusing effect of the errors in the MSSM
can be readily understood by examining the one--loop
solution for $\alpha_3(t)$ in the MSSM.
In the E$_6$SSM the strong gauge coupling 
has a zero one--loop beta function whereas at two--loop level 
the coupling has a mild growth as the
renormalisation scale increases. This implies that 
the uncertainty in the high energy value of $\alpha_3(t)$
in the E$_6$SSM is thus approximately equal to the low 
energy uncertainty in $\alpha_3(t)$.
The relatively large
uncertainty in $\alpha_3(M_X)$ in the E$_6$SSM,
compared to the MSSM, allows one to
achieve exact unification of gauge couplings even within
$1\,\sigma $ deviation of $\alpha_3(M_Z)$ from its average value.

The RG flow of gauge couplings in the E$_6$SSM can be also analysed using the analytical approach 
 for $g_i(t)$ presented in section 2. Substituting the derived two--loop beta 
functions from Eqs.~(\ref{gc10})--(\ref{gc11}) into Eqs.~(\ref{gc1})--(\ref{gc2}) we find the approximate solution
for the $SU(3)_C$, $SU(2)_W$ and $U(1)_Y$ gauge couplings in the E$_6$SSM. The effective threshold scales $\tilde{T}_1$,
$\tilde{T}_2$ and $\tilde{T}_3$ in such a model can be expressed in terms of the MSSM ones, i.e.:
\be
\ba{c}
\tilde{T}_1=T_1^{5/11}m_{H_{\alpha}}^{4/55}\mu_{\tilde{H}_{\alpha}}^{8/55}m_{\tilde{D}_i}^{4/55}\mu_{D_i}^{8/55}
m_{H'}^{2/55}\mu_{\tilde{H}'}^{4/55}\,,\\[4mm]
\tilde{T}_2=T_2^{25/43}m_{H_{\alpha}}^{4/43}\mu_{\tilde{H}_{\alpha}}^{8/43}m_{H'}^{2/43}\mu_{\tilde{H}'}^{4/43}\,,
\qquad\qquad \tilde{T}_3=T_3^{4/7}m_{\tilde{D}_i}^{1/7}\mu_{D_i}^{2/7}\,,
\ea
\label{gc12}
\ee
where $\mu_{D_i}$ and $m_{\tilde{D}_i}$ are the masses of exotic quarks and their superpartners, $m_{H_{\alpha}}$ and 
$\mu_{\tilde{H}_{\alpha}}$ are the masses of non--Higgs and non--Higgsino fields  of the first and second generation,
while $m_{H'}$ and $\mu_{\tilde{H}'}$ are the masses of the scalar and fermion components of $H'$ and $\overline{H}'$. 

The value of strong gauge coupling at the EW scale that results in the exact gauge coupling 
unification can be predicted anew. It is given by Eq.~(\ref{gc4}) where the E$_6$SSM beta functions and new threshold scales 
$\tilde{T}_i$ should be substituted. Such replacement does not change the form of Eq.~(\ref{gc4}) because extra matter 
in the E$_6$SSM form complete $SU(5)$ representations which contribute equally to the one--loop beta functions of the 
$SU(3)_C$, $SU(2)_W$ and $U(1)_Y$ interactions. Due to this the differences of the coefficients of the one--loop beta 
functions $b_i-b_j$ remain intact. But the contributions of two--loop corrections to $\alpha_i(M_X)$ ($\Theta_i$) 
and $\alpha_3(M_Z)$ ($\Theta_s$) change. From Tab.~\ref{gc} it becomes clear that the absolute value
of $\Theta_s$ is considerably smaller in the E$_6$SSM than in the MSSM while the $\Theta_i$'s are a few times larger 
in the former than in the latter. One can also see that the corresponding two--loop corrections 
depend rather weakly on the values of the Yukawa couplings and are almost independent of the extra $U(1)_N$ gauge 
coupling. The dominant contribution to these corrections give $SU(2)_W$ and $SU(3)_C$ gauge couplings which are
considerably larger in the E$_6$SSM as compared with the MSSM case. This explains the large difference between the 
contributions of two--loop corrections to $\alpha_i(M_X)$ in the E$_6$SSM and MSSM. Conversely this is
also a reason why one may expect that the absolute value of the corresponding correction to $\alpha_3(M_Z)$ 
should be at least twice larger in the exceptional SUSY model than in the minimal one leading to the greater
value of $\alpha_3(M_Z)$ at which exact gauge coupling unification takes place.
But due to the remarkable cancellation of different two--loop corrections in Eq.~(\ref{gc4}), the absolute value
of $\Theta_s$ is more than three times smaller in the E$_6$SSM as compared with the MSSM (see Tab.~\ref{gc}). 
Such cancellation is caused by the structure of the two--loop corrections to the beta functions of the SM gauge couplings 
in the considered model. Because of the cancellation of two--loop contributions in Eq.~(\ref{gc4}) the prediction for 
$\alpha_3(M_Z)$ obtained in the  E$_6$SSM is considerably lower than in the MSSM. It is quite 
close to the central value of the recent fit of experimental data even without the inclusion of threshold corrections,
i.e. $\alpha_3(M_Z)\simeq 0.121$.

As in the MSSM the inclusion of threshold effects lowers the prediction for the value of the strong gauge coupling at 
the EW scale. The contribution of threshold corrections $\tilde{\Delta}_s$ to the value of $\alpha_3(M_Z)$, 
that results in the exact gauge coupling unification in the E$_6$SSM, can be parametrised in a manner which is very 
similar to what we had in the MSSM, i.e.
\be
\tilde{\Delta}_s=-\frac{19}{28\pi}\ln\frac{\tilde{M}_{S}}{M_Z}\,,\qquad\qquad 
\tilde{M}_S=\ds\frac{\tilde{T}_2^{172/19}}{\tilde{T}_1^{55/19}\tilde{T}_3^{98/19}}\,.
\label{gc14}
\ee
In the limit when all squarks are degenerate, all sleptons are degenerate, all exotic quarks have the same mass
$\mu_{D_i}$ and the masses of exotic squarks are universal ($m_{\tilde{D}_i}$) we find
\be
\ba{c}
\tilde{M}_S=M_S\cdot\ds\frac{m_{H_{\alpha}}^{12/19}\mu_{\tilde{H}_{\alpha}}^{24/19}\mu^{'18/19}}
{m_{\tilde{D}_i}^{18/19}\mu_{D_i}^{36/19}}=
\mu'\cdot \biggl(\ds\frac{\mu}{\mu'}\biggr)^{1/19}
\biggl(\ds\frac{m_{H_{\alpha}}^{12/19}\mu_{\tilde{H}_{\alpha}}^{24/19}\mu^{18/19}}{m_{\tilde{D}_i}^{18/19} 
\mu_{D_i}^{36/19}}\biggr)\times\\[5mm]
\times\biggl(\ds\frac{m_A}{\mu}\biggr)^{3/19}\biggl(\ds\frac{M_2}{\mu}\biggr)^{4/19}
\biggl(\frac{M_2}{M_3}\biggr)^{28/19}
\biggl(\ds\frac{m_{\tilde{l}}}{m_{\tilde{q}}}\biggr)^{3/19}\simeq 
\mu'\cdot \biggl(\ds\frac{M_{weak}}{M_{colour}}\biggr)^{4.5}\,.
\ea
\label{gc15} 
\ee
Here we also assume that non--Higgs fields of the first two generations have the same mass $m_{H_{\alpha}}$
and the masses of non--Higgsinos of the first and second generation are equal to $\mu_{\tilde{H}_{\alpha}}$
while the masses of scalar non--Higgs fields and their superpartners from $H'$ and $\overline{H}'$ are 
degenerate around $\mu'$. In Fig.~1c--d we keep the masses of all extra exotic particles appearing in the E$_6$SSM 
to be degenerate around $1.5\,\mbox{TeV}$. It means that $\tilde{M}_S=M_S$ in this particular case.
However from Eq.~(\ref{gc15}) it is obvious that in contrast with the MSSM the effective threshold 
scale in the E$_6$SSM is set by $\mu'$. The term $\mu'H'\overline{H}'$ in the superpotential is not involved 
in the process of EW symmetry breaking. Therefore the parameter $\mu'$ remains arbitrary. Because the 
corresponding mass term is not suppressed by the $E_6$ symmetry the components of the doublet superfields $H'$ and 
$\overline{H}'$ are expected to be heavy $\gtrsim 10\,\mbox{TeV}$. As a consequence, although the effective threshold 
scale $\tilde{M}_S$ may be considerably less than $\mu'$, the corresponding mass parameter can be
always chosen so that $\tilde{M}_S$ lies in a few hundred GeV range that allows to get the exact unification
of gauge couplings for any value of $\alpha_3(M_Z)$ which is in agreement with current data.

\section{Conclusions}

In this paper we have presented the two--loop RGEs of the E$_6$SSM and
examined gauge coupling unification in this model using both
analytical and numerical techniques. We have seen that the running
of the gauge couplings in the MSSM and E$_6$SSM are completely
different. For example, in the E$_6$SSM, the strong gauge
coupling grows with increasing renormalisation scale whereas in the
MSSM it decreases at high energies.  Therefore the interval of
variation of $\alpha_3(M_X)$ caused by the uncertainty in
$\alpha_3(M_Z)$ is considerably wider in the E$_6$SSM than in the
MSSM.  Because at any intermediate scale the gauge couplings in the
E$_6$SSM are considerably larger, as compared to the
ones in the MSSM, the two--loop corrections to $\alpha_i(M_X)$ are a
few times bigger in the former than in the latter. At the same time
the absolute value of the corresponding corrections to $\alpha_3(M_Z)$
at which exact gauge coupling unification takes place are much smaller
in the E$_6$SSM than in the MSSM, as is demonstrated in
Table~\ref{gc}. As a consequence the unification of
gauge couplings in the E$_6$SSM can be achieved for
significantly lower values of $\alpha_3(M_Z)$ than in the MSSM. 
The remarkable accidental cancellation of different
two--loop contributions to the prediction for $\alpha_3(M_Z)$ is
caused by the structure of the two--loop corrections to the beta
functions of the gauge couplings in the E$_6$SSM. Thus the
structure of the two--loop contributions to the beta functions of
gauge couplings and large uncertainty in $\alpha_3(M_X)$ allow one to
get exact unification of gauge couplings in the E$_6$SSM for values of
$\alpha_3(M_Z)$ which are within one standard deviation of its
measured central value.

Finally we emphasize that the
effective threshold scale in the E$_6$SSM is set by the mass term of
$H'$ and $\overline{H}'$ from the extra $27'$ and $\overline{27}'$,
which can in principle be very large, significantly 
enhancing the contribution of threshold
corrections to the predictions for $\alpha_3(M_Z)$. 
Indeed, since the only purpose of the 
states $H'$ and $\overline{H}'$ in this model is to achieve
gauge coupling unification, their mass term can be arbitrarily adjusted
to give exact gauge coupling unification in the E$_6$SSM
for any phenomenologically reasonable value of $\alpha_3(M_Z)$.
Future experimental measurements of the mass of these states,
together with more accurate determinations of $\alpha_3(M_Z)$,
will test the self-consistency of this framework.

\section*{Acknowledgements}

\vspace{-2mm} RN would like to thank D.~J.~Miller, G.~G.~Ross, S.~Sarkar, J.~March-Russell, B.~C.~Allanach
and M.~I.~Vysotsky for fruitful discussions. The authors acknowledge support from the SHEFC grant
HR03020 SUPA 36878, the PPARC grant PPA/G/S/2003/00096 and the NATO grant PST.CLG.980066.

\renewcommand{\baselinestretch}{1.00}

\begin{table}[gc]
  \centering
  \begin{tabular}{|c|c|c|c|c|}
    \hline
            & $\Theta_1$ & $\Theta_2$  & $\Theta_3$ & $\Theta_s$ \\
 \hline
MSSM        & $0.556$    & $0.953$     & $0.473$    & $-0.764$ \\
 \hline
E$_6$SSM I  & $1.558$    & $2.322$     & $2.618$    & $-0.250$ \\
 \hline
E$_6$SSM II & $1.604$    & $2.385$     & $2.638$    & $-0.305$ \\
\hline
E$_6$SSM III& $1.602$    & $2.389$     & $2.627$    & $-0.326$ \\
\hline
  \end{tabular}
  \caption{The corrections to $1/\alpha_i(M_X)$ and $1/\alpha_3(M_Z)$
induced by the two--loop contributions to the beta functions in the
MSSM and E$_6$SSM for $\alpha(M_Z)=1/127.9$, $\sin^2\theta_W=0.231$,
$\alpha_3(M_Z)=0.118$ and $\tan\beta=10$.  In the case of the E$_6$SSM
we consider three cases: the scenario E$_6$SSM I corresponds to
$\kappa(M_Z')=\kappa_1(M_Z')=\kappa_2(M_Z')=\lambda(M_{Z'})=
\lambda_1(M_{Z'})=\lambda_2(M_{Z'})=g^{'}_1(M_{Z'})$,
$g^{'2}_1(M_{Z'})=0.227$, $g_{11}(M_{Z'})=0.0202$; in the scenario
E$_6$SSM II we fix $\kappa_i=\lambda_i=0$, $g^{'2}_1(M_{Z'})=0.227$,
$g_{11}(M_{Z'})=0.0202$; in the scenario E$_6$SSM III we ignore all
Yukawa and $U(1)_N$ gauge couplings. Note that in all 
versions of the E$_6$SSM the large individual 
contributions $\Theta_i$ conspire to partially cancel
when forming the quantity $\Theta_s$ which describes
the effect of the two-loop corrections to 
determining the low energy value of $\alpha(M_Z)$.}
  \label{gc}
\end{table}

%\newpage
\begin{figure}
\hspace*{-0cm}{$\alpha_i(t)$}\hspace*{7.7cm}{$\alpha_i(t)$}\\[1mm]
\includegraphics[height=52mm,keepaspectratio=true]{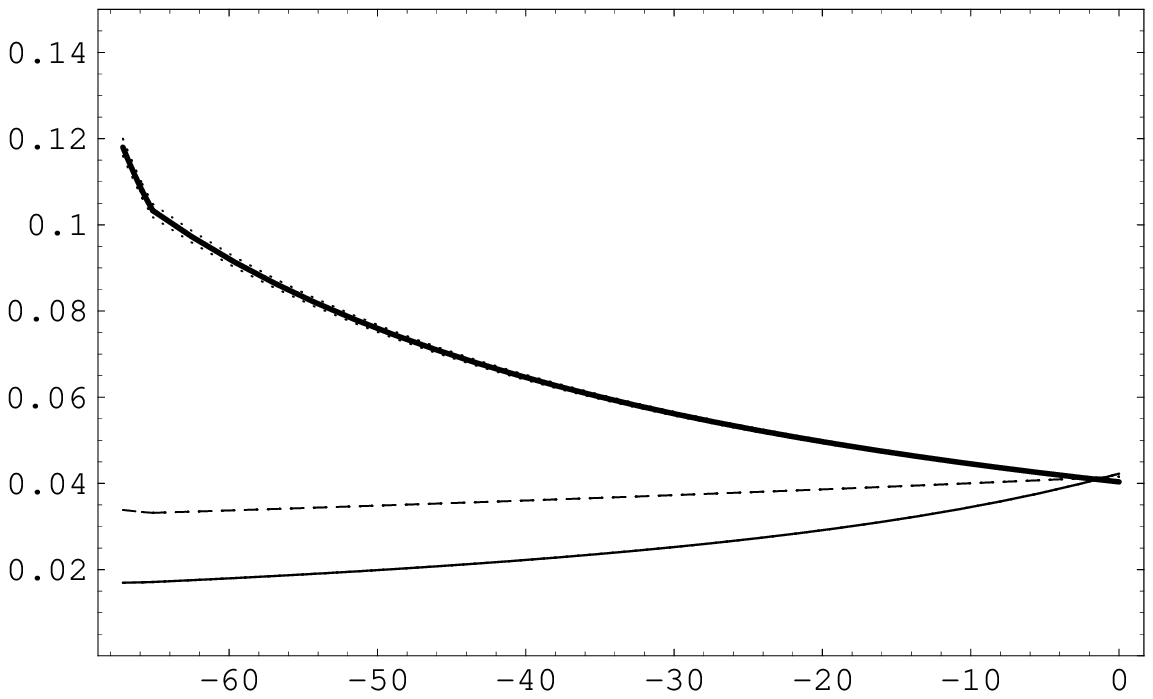}
\includegraphics[height=52mm,keepaspectratio=true]{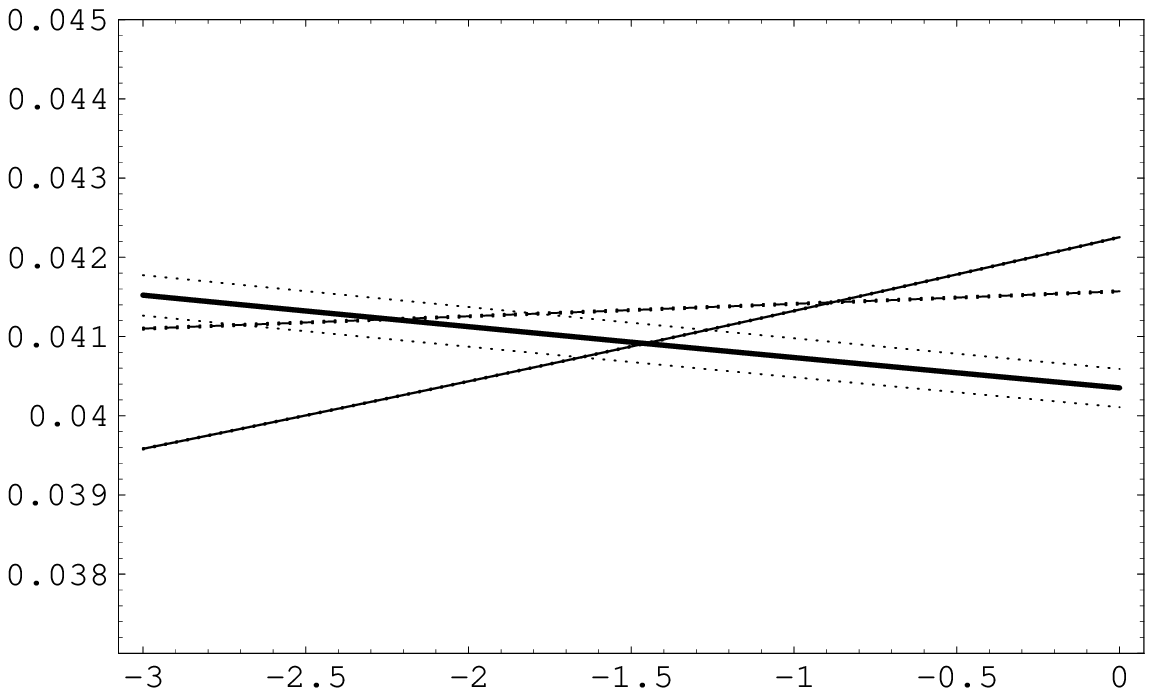}\\
\hspace*{4cm}{$2\log[q/M_X]$}\hspace*{6cm}{$2\log[q/M_X]$}\\[1mm]
\hspace*{5cm}{\bf (a)}\hspace*{7cm}{\bf (b) }\\[3mm]
\hspace*{-0cm}{$\alpha_i(t)$}\hspace*{7.7cm}{$\alpha_i(t)$}\\[1mm]
\includegraphics[height=52mm,keepaspectratio=true]{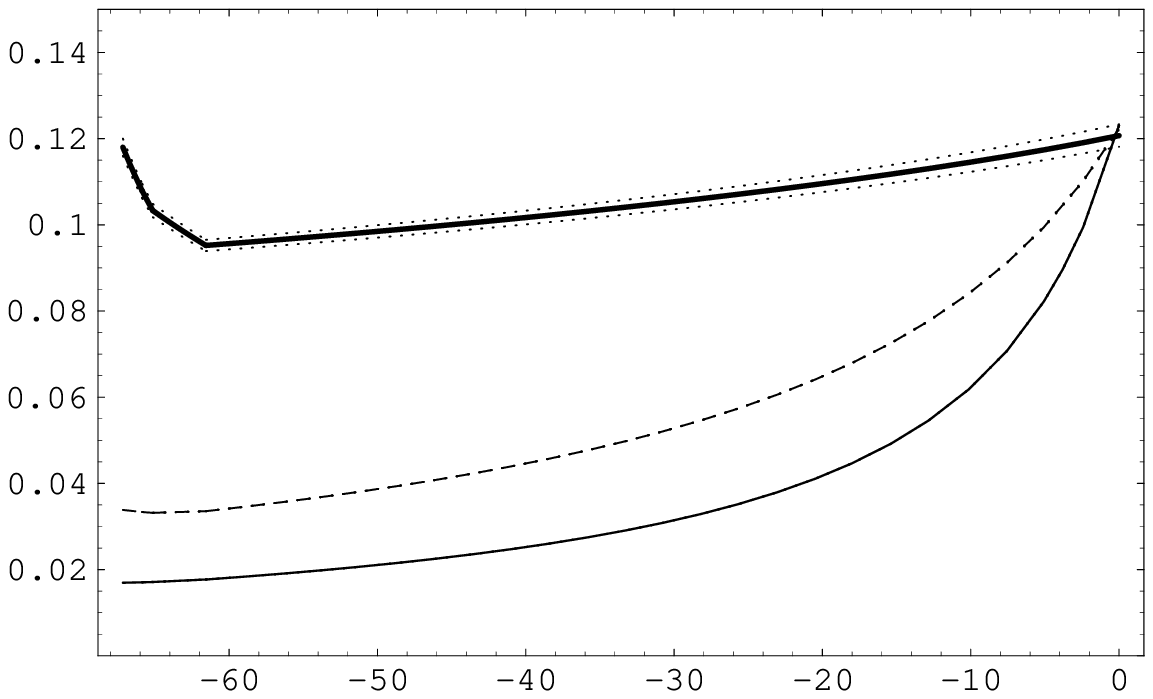}
\includegraphics[height=52mm,keepaspectratio=true]{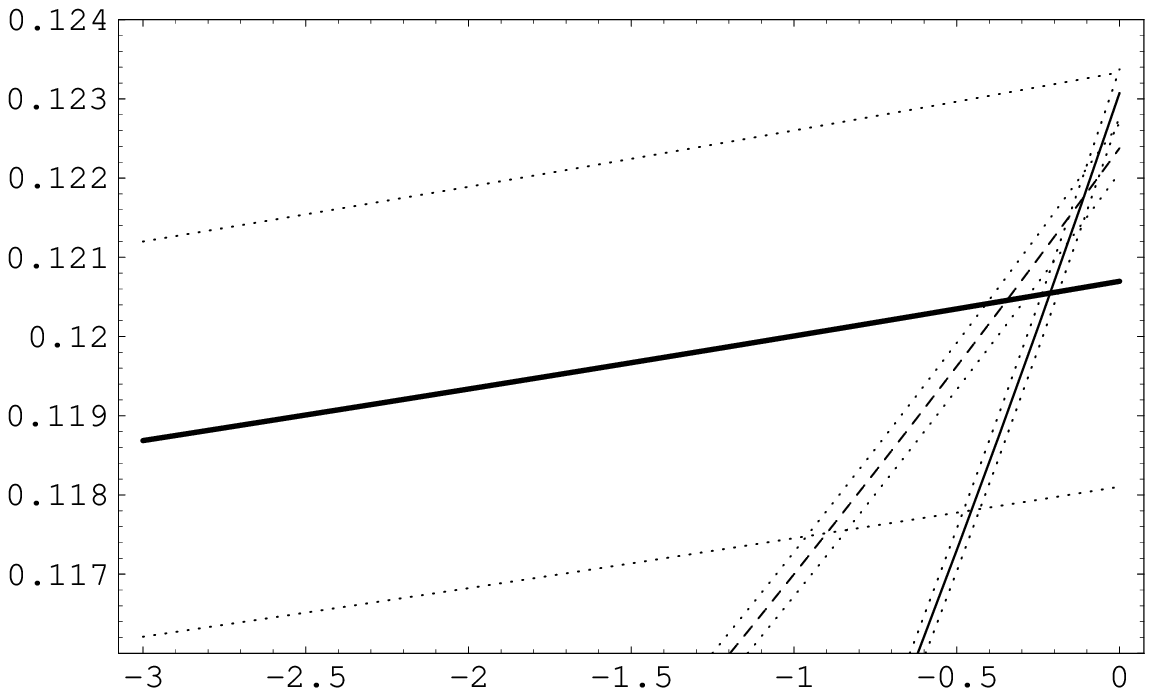}\\
\hspace*{4cm}{$2\log[q/M_X]$}\hspace*{6cm}{$2\log[q/M_X]$}\\[1mm]
\hspace*{5cm}{\bf (c)}\hspace*{7cm}{\bf (d) }\\[3mm]
\caption{Two--loop RG flow of gauge
couplings: {\it (a)} evolution of $SU(3)_C$, $SU(2)_W$
and $U(1)_Y$ couplings from EW to GUT scale $M_X$ 
in the MSSM; {\it (b)} running of SM gauge couplings near the scale 
$M_X$ in the MSSM; {\it (c)} RG flow of $SU(3)_C$, $SU(2)_W$
and $U(1)_Y$ couplings from $M_Z$ to $M_X$
in the E$_6$SSM; {\it (d)} running of SM gauge couplings 
in the vicinity of $M_X$ in the E$_6$SSM. Thick, dashed and
solid lines correspond to the running of $SU(3)_C$, $SU(2)_W$
and~ $U(1)_Y$~ couplings~ respectively.~ We used  $\tan\beta=10$,~ an effective~ SUSY~ 
threshold~ scale~ $M_S=250\,\mbox{GeV}$, $M_{Z'}=1.5\,\mbox{TeV}$,
$\kappa(M_Z')=\kappa_1(M_Z')=\kappa_2(M_Z')=\lambda(M_{Z'})=
\lambda_1(M_{Z'})=\lambda_2(M_{Z'})=g^{'}_1(M_{Z'})$,
$g^{'2}_1(M_{Z'})=0.2271$, $g_{11}(M_{Z'})=0.02024$,
$\alpha_s(M_Z)=0.118$, $\alpha(M_Z)=1/127.9$ and
$\sin^2\theta_W=0.231$. The dotted lines represent the uncertainty 
in $\alpha_i(t)$ caused by the variation of the strong gauge coupling
from 0.116 to 0.120 at the EW scale.}
\label{essmfig1}
\end{figure}


\begin{thebibliography}{99}
\bibitem{1} J. Ellis, S. Kelley, D.V. Nanopoulos, Phys. Lett. B {\bf 260} (1991) 131;
P. Langacker, M. Luo, Phys. Rev. D {\bf 44} (1991) 817;
U. Amaldi, W. de Boer, H. Furstenau, Phys. Lett. B {\bf 260} (1991) 447;
F. Anselmo, L. Cifarelli, A. Peterman, A. Zichichi, Nuovo Cimento {\bf 104}A (1991) 1817, {\bf 105}A (1992) 581.
\bibitem{21} P. Langacker, N. Polonsky, Phys. Rev. D {\bf 52} (1995) 3081.
\bibitem{Chankowski:1995dm} P.H. Chankowski, Z. Pluciennik, S. Pokorski, C.E. Vayonakis,
%``Gauge coupling unification in GUT and string models,''
Phys. Lett. B {\bf 358} (1995) 264.
%[arXiv:hep-ph/9506393].
%%CITATION = HEP-PH 9506393;%%
\bibitem{2} J. Bagger, K. Matchev, D. Pierce, Phys. Lett. B {\bf 348} (1995) 443.
\bibitem{3} W.-M. Yao et al., J. Phys. G {\bf 33} (2006) 1.
\bibitem{31} W. de Boer and C. Sander, Phys. Lett. B {\bf 585} (2004) 276;
W. de Boer, C. Sander, V. Zhukov, A. V. Gladyshev and D. I. Kazakov,
Phys. Lett. B {\bf 636} (2006) 13.
\bibitem{4} S.F. King, S. Moretti, R. Nevzorov, Phys. Rev. D {\bf 73} (2006) 035009;
S.F. King, S. Moretti, R. Nevzorov, hep-ph/0610002. 
\bibitem{5} S.F. King, S. Moretti, R. Nevzorov, Phys. Lett. B {\bf 634} (2006) 278. 
\bibitem{6} D.R.T. Jones, Nucl. Phys. B {\bf 87} (1975) 127;
K. Inoue, A. Kakuto, S. Takeshita, Prog. Theor. Phys. {\bf 67} (1982) 1889; ibid. {\bf 68} (1982) 927;
D.R.T. Jones, L. Mezincescu, Phys. Lett. B {\bf 136} (1984) 242;
P. West, Phys. Lett. B {\bf 137} (1984) 371;
A. Parkes, P. West, Phys. Lett. B {\bf 138} (1984) 99;
D.R.T. Jones, L. Mezincescu, Phys. Lett. B {\bf 138} (1984) 293;
M.E. Machacek, M.T. Vaughn, Nucl. Phys. B {\bf 236} (1984) 221.
\bibitem{7} S.P. Martin, M.T. Vaughn, Phys. Rev. D {\bf 50} (1994) 2282.
\bibitem{8} I. Antoniadis, C. Kounnas, K. Tamvakis, Phys. Lett. B {\bf 119} (1982) 377;
I. Antoniadis, C. Kounnas, R. Lacaze, Nucl. Phys. B {\bf 211} (1983) 216. 
\bibitem{Carena:1993ag} M. Carena, S. Pokorski, C.E.M. Wagner,
%``On the unification of couplings in the minimal supersymmetric Standard
%Model,''
Nucl. Phys. B {\bf 406} (1993) 59.
\bibitem{Langacker:1992rq} P. Langacker, N. Polonsky,
%``Uncertainties in coupling constant unification,''
Phys. Rev. D {\bf 47} (1993) 4028. 
\bibitem{9} G.G. Ross, R.G. Roberts,
%``Minimal supersymmetric unification predictions,''
Nucl. Phys. B {\bf 377} (1992) 571;
V.D. Barger, M.S. Berger, P. Ohmann,
%``Supersymmetric grand unified theories: Two loop evolution of gauge and
%Yukawa couplings,''
Phys. Rev. D {\bf 47} (1993) 1093;
P. Langacker, N. Polonsky,
%``The Bottom mass prediction in supersymmetric grand unification:
%Uncertainties and constraints,''
Phys.\ Rev.\ D {\bf 49} (1994) 1454.
\bibitem{10} E. Keith, E. Ma, Phys. Rev. D {\bf 56} (1997) 7155.
\bibitem{11} S. Kraml {\it et al.} (eds.), {\it Workshop on CP studies and 
non-standard Higgs physics}, CERN--2006--009, hep-ph/0608079;
S.F. King, S. Moretti, R. Nevzorov, hep-ph/0601269.
\bibitem{12} J. Rich, M. Spiro, J. Lloyd-Owen, Phys. Rept. {\bf 151} (1987) 239;
P.F. Smith, Contemp. Phys. {\bf 29} (1988) 159; T.K. Hemmick {\it et al.},
Phys. Rev. D {\bf 41} (1990) 2074.
\bibitem{121} Y.~Kawamura,
%``Triplet-doublet splitting, proton stability and extra dimension,''
Prog.\ Theor.\ Phys.\  {\bf 105} (2001) 999;
G.~Altarelli and F.~Feruglio,
%``SU(5) grand unification in extra dimensions and proton decay,''
Phys.\ Lett.\  B {\bf 511} (2001) 257;
L.~J.~Hall and Y.~Nomura,
%``Gauge unification in higher dimensions,''
Phys.\ Rev.\  D {\bf 64} (2001) 055003;
A.~Hebecker and J.~March-Russell,
%``A minimal S(1)/(Z(2) x Z'(2)) orbifold GUT,''
Nucl.\ Phys.\  B {\bf 613} (2001) 3;
T.~Asaka, W.~Buchmuller and L.~Covi,
%``Gauge unification in six dimensions,''
Phys.\ Lett.\  B {\bf 523} (2001) 199;
L.~J.~Hall, Y.~Nomura, T.~Okui and D.~R.~Smith,
%``SO(10) unified theories in six dimensions,''
Phys.\ Rev.\  D {\bf 65} (2002) 035008.
\bibitem{13} K. S. Babu, C. Kolda, J. March--Russell, Phys. Rev. D {\bf 54} (1996) 4635;
P. Langacker, J. Wang, Phys. Rev. D {\bf 58} (1998) 115010;
D. Suematsu, Phys. Rev.  D {\bf 59} (1999) 055017; 
T. G. Rizzo, Phys. Rev. D {\bf 59} (1999) 015020. 


\end{thebibliography}
\end{document}